\documentclass[11pt]{article}
\usepackage{amssymb,amsmath}
\include{epsf}

\usepackage{graphicx}
\usepackage{amsfonts}

\newcommand{\e}{{\mathrm e}}
\newcommand{\ii}{{\mathrm i}}
\newcommand{\dd}{{\mathrm d}}
\newcommand{\eqn}[1]{(\ref{#1})}

\def\appendix#1{\addtocounter{section}{1}\setcounter{equation}{0}
\renewcommand{\thesection}{\Alph{section}}
\section*{
\thesection\protect\indent \parbox[t]{11.715cm} {#1}}
\addcontentsline{toc}{section}{Appendix\thesection\ \ \ #1} }
\newcommand{\real}{{\mathbb R}} 
\newcommand{\R}{{\mathbb R}} 

\newcommand{\tr}[1]{\:{\rm tr}\,#1}
\newcommand{\Tr}[1]{\:{\rm Tr}\,#1}
\newcommand{\be}{\begin{equation}}
\newcommand{\ee}{\end{equation}}
\newcommand{\beq}{\begin{equation}}
\newcommand{\eeq}{\end{equation}}
\newcommand{\bea}{\begin{eqnarray}}
\newcommand{\eea}{\end{eqnarray}}
\def\beqa{\begin{eqnarray}}
\def\eeqa{\end{eqnarray}}

\newcommand{\del}{\partial}

\newcommand{\eq}{\begin{equation}}
\newcommand{\eqa}{\begin{eqnarray}}
\newcommand{\en}{\end{equation}}
\newcommand{\ena}{\end{eqnarray}}

\newcommand{\starmoy}{\star_M}

\begin{document}
\begin{titlepage}


\begin{center}

\baselineskip=24pt

{\Large\bf On the  $\star$-product quantization and the Duflo map in  three dimensions}

\baselineskip=14pt

\vspace{1cm}

{ Luigi Rosa and Patrizia Vitale }
\\[6mm]
{\it Dipartimento di Scienze Fisiche, Universit\`{a} di Napoli {\sl Federico II} }\\
	{and}\\
{ \it INFN, Sezione di Napoli}\\
{\it Monte S.~Angelo, Via Cinthia, 80126 Napoli, Italy}
\\[4mm]
{\small\tt
luigi.rosa@na.infn.it, patrizia.vitale@na.infn.it}

\end{center}

\vskip 2 cm

\begin{abstract}
We analyze the $\star$-product induced on $\mathcal{F}(\mathbb{R}^3)$  by a suitable reduction of the Moyal product defined on
$\mathcal{F}(\mathbb{R}^4)$. This is obtained through the identification $\mathbb{R}^3\simeq\mathfrak{g}^*$,  with $\mathfrak{g}$ a  three-dimensional Lie algebra.
 We consider the $\mathfrak{su}(2)$ case,  exhibit a matrix basis and realize the  algebra of functions on
$\mathfrak{su}(2)^*$ in such a basis. The relation to the Duflo map is discussed. As an application to quantum mechanics we compute the spectrum of the hydrogen atom.
\end{abstract}

\end{titlepage}

\section*{Introduction}
There has been recently a renewed interest in noncommutative structures on
$\mathbb{R}^3$, mainly because of their occurrence in quantum gravity models
\cite{qgpeople}, where $\mathbb{R}^3$, or copies of it, are identified with the dual algebra of the
appropriate Lorentz group. In such framework  $ \star$ products are mainly introduced
through a group Fourier transform (see for example \cite{Majid,noui}), invoking
compatibility with the group convolution. However, direct definitions are available in
the literature, which don't rely on the existence of a group Fourier transform and have
obvious calculational advantages. In view of applications to quantum field theory
(\cite{vitalewallet}) we concentrate on this approach and refer to the cited literature
for the other (cfr. \cite{PV12} where a comparison is made in the last section).

Besides their appearence in the quantum gravity context, noncommutative structures on $\R^3$ are very interesting because of their emergence in the quantization of standard dynamical systems.
To this, we shall discuss in some detail the application to the quantization of the hydrogen atom. As for their implications in quantum field theory, such that their  renormalization properties and the UV/IR behaviour, which are quite different from noncommutative field theories on Moyal (hyper-)planes,  they are analysed in   \cite{vitalewallet}.

The main object of this paper is to study in detail a family of star products introduced long ago in \cite{selene}. These products are obtained by a suitable reduction of the Moyal product on $\R^4$ on identifying the coordinate functions on $\R^3$ with one of the many subalgebras of the algebra of quadratic-linear functions on $\R^4$, the latter being isomorphic to the symplectic algebra $\mathfrak{sp}(4)$. This procedure induces on $\R^3$ a non-constant noncommutativity of Lie algebra type. We concentrate on the $\mathfrak{su}(2)$ case, introduce a matrix basis and, in the spirit of deformation quantization, propose an application to the quantization of the hydrogen atom. At the same time we review a mathematically preferred quantization map, the Duflo isomorphism and show that our $\star$-product and the related quantization scheme are singled out among other possible definitions in that they agree with the Duflo map.

The paper is organised as follows. In Section \ref{sec.1} we review the family
of $\star$ products introduced in \cite{selene} and   we focus  on the $SU(2)$ related
$\star$-product. In Section \ref{sec.2}. we define  a matrix basis through a suitable reduction of the Moyal
(hyper)-plane matrix basis \cite{pepejoe}, which has been so crucial for the
renormalizability proof of the Grosse-Wulkenhaar model \cite{grossewulkenhaar}. The exhistence of a matrix basis for the three dimensional case is an important calculational tool, as it reduces the   $\star$ product to a matrix product.
In Section \ref{sec.3} after a short revision of  the Duflo isomorphism we compute its value for   the quadratic Casimir of $SU(2)$
and apply it to the quantization of the hydrogen atom, showing that it reproduces the correct spectrum if the whole symmetry of the system is taken into account. In Section \ref{sec.4} we perform the same calculation within the  $\star$-quantization
associated to the $\star$ product defined in Sections \ref{sec.1} and \ref{sec.2} and show that this $\star$ product is singled out among other possible definitions because of its agreement with the Duflo quantization. We conclude with some comments and perspectives.

\section{Moyal induced star products on $\mathcal{F}(\mathbb{R}^3)$}\label{sec.1}

Let us   consider the Moyal product  \cite{Gronewold,Moyal}  defined on on $\mathcal{F}(\mathbb{R}^4)\simeq \mathcal{F}(\mathbb{C}^2)$.  On using complex
coordinates
$ (z^a, \bar z^a), a=1,2$  we have
\be (
f  \starmoy g)(z,\bar z)=f(z,\bar z)
\exp \left[\frac{\theta}{2}(\overleftarrow{\del_z^a}\overrightarrow{\del_{\bar z^a}}
-\overleftarrow{\del_{\bar z^a}}
  \overrightarrow{\del_z^a})\right] g(z,\bar z)
\label{moy}
\ee
where the operator $\overleftarrow{\del}$ (resp.\ $\overrightarrow{\del}$) acts on the
left (resp.\ on the right). It is well known that this product   is induced through the
Weyl quantization map \cite{Weyl} which, in two dimensions,  associates
 to a function on the plane an operator according to:
\be
\hat f=\hat {\mathcal{W}}(f)=\frac{1}{(2\pi)^2}\int\dd^2 z \,\hat \Omega (z, \bar z)
f(z, \bar z) \label{Weylmap}
\ee
where
\be
\hat \Omega (z, \bar z)= \int \dd^2 \eta\, e^{{ -( \eta \bar z-\bar\eta z )}} \e^{
\theta( \eta a^\dag-\bar\eta a )}
\ee
is the so called quantizer and $a, a^\dag$ are the usual (configuration space) creation
and annihilation operators, with commutation relations
\be
[a, a^\dag]=\theta. \label{commz}
\ee
The inverse map (the Wigner map \cite{Wigner})  is represented by:
\be
f(z,\bar z) = \mathcal{W}^{-1}(\hat f)= \Tr [\hat f \hat \Gamma(z, \bar z)] \label{wigner}
\ee
with $\hat \Gamma(z,\bar z) = \hat\Omega (z,\bar z)$, meaning that the Weyl-Wigner
quantization/dequantization procedure  is self-dual (see \cite{pepejoe,conVolodya} for
details).

 The Moyal product is then defined as
\be
f\starmoy g = \mathcal{W}^{-1}\left(\hat{\mathcal{W}}(f)\hat {\mathcal{W}}(g)\right).
\label{moyal}
\ee

 The Weyl-Wigner-Moyal quantization scheme is the prototype of deformation
quantization \cite{bayen}, where the algebra of operators is replaced by  an algebra of
functions with noncommutative product. From the latter expression it is not difficult (see
for example~\cite{zampinithesis}) to obtain integral expressions for the product, a few
of which are collected in~\cite[Appendix]{selene}. The standard expression \eqn{moy} is
indeed an asymptotic expansion of the integral expressions~\cite{VarillyGraciaBondia}.
It results
\bea
z {\starmoy}  \bar z&=& z\cdot \bar z +\frac{\theta}{2}\nonumber\\
\bar z {\starmoy}  z&=& z\cdot \bar z -\frac{\theta}{2}. \label{starcommz}
\eea
The dimension of the parameter $\theta$ depends on the physical meaning of the carrier space $\R^2$. In what follows we will choose the coordinate functions $z_a,\bar z_a$ and $\theta$ to be dimensionless. The generalization to the four dimensional case is straightforward.  We recall that  Weyl
quantization associates to polynomial functions operators in the symmetric ordering,  as opposed to other quantization
schemes (see for example \cite{voros}).

What is properly defined as the Moyal algebra is $\mathcal{M}_\theta:=
\mathcal{M}_L({\real}_\theta^4)\cap \mathcal{M}_R({\real}_\theta^4)$ where
$\mathcal{M}_{L}({\real}_\theta^4)$, the left multiplier algebra, is defined as the
subspace of tempered distributions that give rise to Schwartz functions when left
multiplied by Schwartz functions;  the right multiplier
algebra~$\mathcal{M}_R({\real}_\theta^4)$ is analogously defined. For more details we
refer to the appendix in \cite{selene} and references therein. In the present article
we shall think of $\mathcal{M}_\theta$ as the algebra of $\ast$-polynomial functions in
$z^a,\bar z^a$ properly completed. Its commutative limit, $\mathcal{F}(\real^4)$, is the
commutative multiplier algebra $\mathcal{O}_M(\real^4)$, the algebra of smooth functions
of polynomial growth on $\real^4$ in all derivatives \cite{GGISV04}.

The crucial step  to obtain star products on $\mathcal{F}(\mathbb{R}^3)$, hence to deform $\mathcal{F}(\mathbb{R}^3)$ into a noncommutative algebra, is to identify
$\mathbb{R}^3$ with the dual, $\mathfrak{g}^*$,  of some chosen three dimensional Lie
algebra $\mathfrak{g}$. This identification induces on $\mathcal{F}(\mathbb{R}^3)$ the
Kirillov Poisson bracket, which, for coordinate functions reads
\be
\{x_i,x_j\}=c_{ij}^k x_k \label{Kirillov}
\ee
with $i=1,..,3$ and $c_{ij}^k $ the structure constants of $\mathfrak{g}$.  On the other
hand, all three-dimensional (Poisson) Lie algebras may be realized as subalgebras of the
symplectic algebra $\mathfrak{sp}(4)$, which is classically realized  as the Poisson
 algebra of quadratic-linear functions on $\mathbb{R}^4$ ($\mathbb{C}^2$ with
our choices) with canonical Poisson bracket
\be
\{z^a, \bar z^b\}= \ii.
\ee
It is then  possible to find quadratic-linear functions
\be x_i= x_i(z^a,\bar z^a)
\ee
 which obey \eqn{Kirillov}.
This is nothing but the classical counterpart of the Jordan-Schwinger map realization of
Lie algebra generators in terms of creation and annihilation operators \cite{MMVZ94}. Then
one can show \cite{selene} that these Poisson subalgebras are also Moyal subalgebras,
that is
\be
x_i(z^a,\bar z^a)\starmoy x_j(z^a,\bar z^a) -x_j(z^a,\bar z^a)\starmoy x_i(z^a,\bar
z^a)= \lambda c_{ij}^k x_k(z^a,\bar z^a) \label{starx}
\ee
where the noncommutative parameter $\lambda$ depends on  $\theta$ and shall be adjusted according to  the physical dimension of the coordinate functions $x_i$.
Occasionally we shall indicate with $\mathbb{R}^3_\lambda$ the  noncommutative
algebra $(\mathcal{F}(\R^3), \star)$.  Eq. \eqn{starx}  induces a star product on polynomial functions on
$\mathbb{R}^3$ generated by the coordinate functions $x_i$, which may be expressed in
closed form in terms of differential operators on $\mathbb{R}^3$. For details we refer
to \cite{selene} where all products are classified. Here we will consider quadratic
realizations of the kind
\be
\pi^*(x_\mu)=\kappa \bar z^a e_\mu^{ab} z^b, \;\;\; \mu=0, ..,3 \label{xmu}
\ee
with  $e_i= \frac{1}{2}\sigma_i, \: i=1,..,3$ are the $SU(2)$ generators and $\sigma_i$
are the Pauli matrices, while $e_0=\frac{1}{2} \mathbf{1}$. Here we have explicitly indicated the pull-back map $\pi^*:\mathcal{F}(\R^3)\mapsto \mathcal{F}(\R^4)$. We will shall omit it in the following, unless necessary. $\kappa$ is some possibly dimensional constant such that $\lambda=\kappa \theta$.  Notice that
\be
x_0^2= \sum_i x_i^2.
\ee
It is possible to show that \cite{selene}
\be
(x_i\star \phi) (x) = \left\{x_i
-i\frac{\lambda}{2}\epsilon_{ijk}x_j\del_k-\frac{\lambda^2}{8}[(1+x\cdot
\del)\del_i-\frac{1}{2}x_i\del\cdot\del]\right\}\phi(x) \label{starsu2}
\ee
which implies for coordinate functions
\be
x_i\star  x_j= x_i \cdot x_j + i\frac{\lambda}{2}\epsilon_{ijk}x_k
-\frac{\lambda^2}{8}\delta_{ij}
\ee
\be
x_i,\star  x_j-x_j\star x_i=  i{\lambda}\epsilon_{ijk}x_k \label{inducedstarcom}
\ee
 where we have indicated with $\star$ the induced star product.
Moreover we have
\be
[x_0, \phi(x_i)]_\star=0 \label{commutant}
\ee
That is, $x_0$ lies in the center of  the algebra $\mathbb{R}_\lambda^3$, so that the latter  may be equivalently defined as the commutant of $x_0$.

It is possible to use other star products on $\mathbb{R}^4$ such that the polynomial algebra generated by the coordinate functions in Eq. \eqn{xmu} is  still a subalgebra with the same $\star$-commutator. See for example \cite{HLS} where the Voros product has been used.  This corresponds to the choice of normal ordering at the level of $\star$ quantization in $\mathbb{R}^4$. For applications of the Voros and other translation-invariant $\star$ products to 4-dimensional quantum field theory (QFT) see for example \cite{GLV08}, \cite{GLV09}, \cite{TV10}.  While these products  yield the same $\star$ commutator between coordinates, they might have   important consequences  which will be discussed in the last section.

The expression \eqn{starsu2} which was derived in \cite{selene} for the star product in $\R^3_\lambda$ is practically difficult to use when we need to extend it to the product of two generic functions.  In next section we shall derive a matrix basis for $\R^3_\lambda$ which makes it much easier to compute  the $\star$-product. We shall see that it reduces the $\star$ product \eqn{starsu2} to matrix multiplication.

\section{Matrix formulation of  $\mathbb{R}^3_\lambda$}\label{sec.2}

In this section we derive a matrix formulation of $ \mathbb{R}^3_\lambda$  which is based
on a suitable reduction of the matrix formulation of the Moyal space
$\mathbb{R}^4_\theta$ first introduced by Gracia-Bondia and Varilly in \cite{pepejoe}.
We also discuss the meaning of the new matrix basis in connection to the $\star$-quantization of the angular momentum and to the Wigner functions associated to the density operator of angular momentum states.
To this, let us first review  the derivation in \cite{pepejoe} for the two and four dimensional case.
\subsection{The matrix basis of $\mathbb{R}^{2n}_\theta$}
To a function on $\mathbb{R}^2$ we associate via the Weyl map  \eqn{Weylmap} the operator
\be
\phi (z,\bar z)\rightarrow \hat \phi (a,  a^\dag).
\ee
This may be expanded into symmetric ordered powers of $a, a^\dag$
\be
\hat\phi =\sum_{p,q} \tilde\phi_{p q} :{a^\dag}^p  a^q: \label{phi'}
\ee
with $: \; :$ denoting the symmetric ordering. On using the number basis
\be
a^\dag a |n\rangle = n | n\rangle
\ee
together with
\be
a|n\rangle= \sqrt{n\theta}|n-1\rangle\;\;\; a^\dag |n\rangle =
\sqrt{n+1)\theta}|n+1\rangle
\ee
we may rewrite \eqn{phi'} as
\be
\hat\phi =\sum_{p,q} \phi_{p q} |p\rangle\langle q|
\ee
with $\tilde\phi_{pq}, \phi_{k l}$ related by a change of basis
\be
\phi_{lk}=\sum_{q=0}^{min\,(l,k)} \tilde \phi_{l-q,k-q} \frac{\sqrt{l! k! \theta^{l+k}}}{\theta^q q!}\label{changeofbasis}
\ee

 On applying the Wigner map we obtain a function in the
noncommutative Moyal algebra
\be
\phi(z, \bar z) =\sum_{p,q} \phi_{p q} f_{pq}(z, \bar z)
\ee
with
\be
f_{pq}(z, \bar z)= \mathcal{W}^{-1}(|p\rangle\langle q|)=
\frac{1}{\sqrt{p!q!\theta^{pq}}}\bar z^p \starmoy f_{00}\starmoy z^q \label{wigner functions}
\ee
and
\be
f_{00}(\bar z, z)= 2 \exp(-2\bar z z/\theta). \ee
 This Gaussian is  idempotent w.r.t. the Moyal star product as it  satisfies the condition
 \be
 f_{00}\starmoy f_{00} (\bar z, z) = f_{00} (\bar z, z) \label{f00}
 \ee
 Let us recall that, when qualifying $\R^2$  as the phase space of 1-dimensional systems,  the basis functions $f_{pq}(z, \bar z)$ in the $\star$-quantization scheme
correspond exactly to the Wigner functions associated to the density operator of the quantum oscillator states.

The basis elements $f_{pq}(z,\bar z)$ may be seen to obey
\be
f_{pq}\starmoy f_{kl}=\delta_{qk}f_{pl}
\ee
either by direct calculation, on using \eqn{f00}, or by observing that, by definition
\be
f_{pq}\starmoy f_{kl}=\mathcal{W}^{-1}(|p\rangle\langle q|k\rangle\langle l|).
\ee
Moreover
\be
\int \dd^2 z\, f_{pq}(z,\bar z)= 2\pi \theta \delta_{pq}.
\ee
The extension to $\mathbb{R}^4_\theta$ is straightforward. We have
\be
\phi(z_a,\bar z_a)= \sum_{\vec p \vec q} \phi_{\vec p \vec q} f_{\vec p \vec q}(z_a,\bar
z_a) \label{phir4}
\ee
with $a=1,2$, $\vec p= (p_1,p_2)$ and
\be
f_{\vec p \vec q}(z_a,\bar z_a)=\mathcal{W}^{-1}(|p_1 p_2\rangle\langle q_1
q_2|)=f_{p_1,q_1} (z_1,\bar z_1)\cdot f_{p_2,q_2} (z_2,\bar z_2) . \label{fpq4}
\ee
In order to describe elements of $\mathbb{R}_\theta^2$ (resp. $\mathbb{R}_\theta^4$),
the sequences $\{\phi_{pq}\}$ (resp. $\{\phi_{\vec p\vec q}\})$  have to be of rapid
decay \cite{pepejoe}.
\subsection{The matrix basis of $\mathbb{R}^3_\lambda$}\label{subsec.2.2}
In order to obtain a matrix basis in three dimensions, compatible with the product
\eqn{starsu2}, we resort to the Schwinger-Jordan realization of the $\mathfrak{su}(2)$
Lie algebra in terms of creation and annihilation operators. We observe that the
eigenvalues of the number operators $\hat N_1=a^\dag_1 a_1$, $\hat N_2=a^\dag_2 a_2$, say  $p_1,
p_2$, are related to the eigenvalues of $\hat{\mathbf{X}}^2, \hat X_3$, respectively $j(j+1)$ and $m$,   by
\be
p_1+p_2=2j\;\;\; p_1-p_2=2m
\ee
with $p_i\in \mathbb{N}$, $j\in \mathbb{N}/2$, $-j\le m \le j$,
so to have
\be
|p_1 p_2\rangle=|j+m, j-m\rangle\equiv |j m\rangle=
\frac{(a_1^\dag)^{j+m}(a_2)^{j-m}}{\sqrt{(j+m)!(j-m)!}} |00\rangle
\ee
where $\hat X_i, i=1,..,3$ are the standard angular momentum operators  representing the $\mathfrak{su}(2)$ Lie algebra in terms of selfadjoint operators on the Hilbert space spanned by $|j,m\rangle$.
Then we may relabel the matrix basis of $\mathbb{R}^4_\theta$, Eq. \eqn{fpq4} as
$f^{j\tilde \j}_{m \tilde m}$, so to have
\be
\phi(z_a,\bar z_a)= \sum_{j \tilde\j\in \mathbb{N}/2}\sum_{m=-j}^j \sum_{\tilde m=-\tilde\j}^{\tilde\j}
\phi^{j \tilde\j}_{m\tilde m} f^{j\tilde\j}_{m\tilde m}(z_a,\bar z_a)
\ee
We further observe that,  for  $\phi$ to be in the subalgebra
$\mathbb{R}^3_\lambda$ we must impose $j=\tilde\j$. To this it suffices to
compute
\be
x_0 \star f^{j\tilde\j}_{m\tilde m}-f^{j\tilde\j}_{m\tilde m}\star
x_0=\lambda(j-\tilde\j) f^{j\tilde\j}_{m\tilde m}
\ee
and remember that $\mathbb{R}_\lambda^3$  may be alternatively defined as
the $\star$-commutant of $x_0$. This requires
\be
j=\tilde \j
\ee
We have then
\be
\phi(x_i)=\sum_{j}\sum_{m,\tilde m=-j}^j \phi^j_{m\tilde m} v^j_{m\tilde m}
\label{phixi}
\ee
with
\be
v^j_{m\tilde m}:= f^{jj}_{m\tilde m}=\frac{\bar z_1^{j+m}\starmoy f_{00}(\bar z_1,z_1)
\starmoy z_1^{j+\tilde m} \bar z_2^{j-m}\starmoy f_{00}(\bar z_2,z_2) \starmoy
z_2^{j-\tilde m}}{\sqrt{(j+m)!(j-m)! (j+\tilde m)!(j-\tilde m)! \theta^{4j} }} \label{matrixbasis}
\ee
It is useful to recall its expression in terms of the Wigner map. We have
\be
v^j_{m\tilde m}=\mathcal{W}^{-1}(|j\,m\rangle j\,\tilde m|)\label{spinwignerfunctions}
\ee
In view of the application to Quantum Mechanics one can recognize this as  the Wigner functions of the angular momentum eigenstates, that is the Wigner symbols associated to the density operator (cfr. for example \cite{pepespin}).\footnote{ See also \cite{volodyabeppe} where symbols for the angular momentum states are introduced in a different $\star-$quantization scheme, the so called tomographic or probability representation of quantum mechanics \cite{MMT,noiconvolodya}).}
The orthogonality property now reads
\be
v^j_{m\tilde m}\star v^{\tilde\j}_{n \tilde n}=\delta^{j \tilde\j}\delta_{\tilde m
n}v_{m \tilde n}.
\ee
As for the normalization we have
\be
\int \dd^2 z_1\dd^2 z_2 \;v^j_{m \tilde m} (z, \bar z)= 4 \pi^2 \theta^2 \delta_{m
\tilde m}
\ee
 The star
product in $\mathbb{R}_\lambda^3$ becomes a matrix product
\be
\phi\star \psi (x)=\sum \phi^{j_1}_{m_1\tilde m_1} \phi^{j_2}_{m_2\tilde m_2}
v^{j_1}_{m_1\tilde m_1} \star v^{j_2}_{m_2\tilde m_2} =\sum \phi^{j_1}_{m_1\tilde m_1}
\phi^{j_2}_{m_2\tilde m_2} v^{j_1}_{m_1\tilde m_2} \delta^{j^1 j^2} \delta_{\tilde m_1
m_2}
\label{starmatrix}
\ee
while the integral may be defined through the pullback to $\R^4_\theta$
\be
\int_{\mathbb{R}^3_\lambda} \phi \star \psi
:=\kappa^2 \int_{\mathbb{R}^4_\theta}\pi^\star(\phi)\starmoy \pi^*(\psi)= 4\pi^2 \lambda^2 \Tr
\Phi\Psi
\ee
hence becoming a trace.\footnote{
If we were to perform our analysis in the coordinate basis, without recurring to the matrix basis, we  should use a differential calculus adapted to   $\R^3_\lambda$ as the one introduced in  \cite{diffcalc} .}

In analogy with the present derivation, the matrix basis adapted to the Voros product can be found in \cite{discofuzzy}, and it is reduced to 3 dimensions in \cite{vitalewallet} where applications to 3d QFT are studied.

\subsection{The $\star$-quantization of the the $SU(2)$ generators in the matrix basis}\label{angularmomentum}

For further application to the quantization of the hydrogen atom let us work out in detail the realization of the $SU(2)$ generators in the matrix basis.
On using the expression of the generators in terms of $\bar z_a, z_a$ and the basis transformations \eqn{changeofbasis} we find their expression in the matrix basis
\beqa
x_+ &=&\kappa\bar z_1 z_2 = \lambda \sum_{j,m}\sqrt{(j+m)(j-m+1)} v^j_{m\,m-1}\\
x_- &=& \kappa\bar z_2 z_1 = \lambda \sum_{j,m}\sqrt{(j-m)(j+m+1)} v^j_{m\,m+1}\\
x_3 &=&\frac{\kappa}{2} (\bar z_1 z_1-\bar z_2 z_2)=\lambda \sum_{j,m} m v^j_{m\,m}\\
x_0 &=&\frac{\kappa}{2} (\bar z_1 z_1+\bar z_2 z_2)=\lambda \sum_{j,m} (j+\frac{1}{2}) v^j_{m\,m}
\eeqa
were we have introduced
\be
x_\pm:= x_1\pm i x_2
\ee
This yields
\be
\begin{array}{lll}
x_+\star v^j_{m\tilde m}= \lambda \sqrt{(j+m+1)(j-m)} v^j_{m+1 \, \tilde m}  & & v^j_{m\tilde m}\star x_+= \lambda\sqrt{(j-\tilde m+1)(j+\tilde m)} v^j_{m\, \tilde m -1} \\
x_-\star v^j_{m\tilde m}= \lambda \sqrt{(j-m+1)(j+m)} v^j_{m-1\,\tilde m}  &&v^j_{m\tilde m}\star x_-= \lambda\sqrt{(j+\tilde m+1)(j-\tilde m)} v^j_{m \,\tilde m +1} \\
x_3\star v^j_{m\tilde m}= \lambda\, m \, v^j_{m\tilde m}&&  v^j_{m\tilde m} \star x_3= \lambda \,\tilde m \, v^j_{m\tilde m}\\
x_0\star v^j_{m\tilde m}= \lambda( j+\frac{1}{2}) \, v^j_{m\tilde m}&&  v^j_{m\tilde m} \star x_0= \lambda ( j+\frac{1}{2}) \,v^j_{m\tilde m}
\end{array}
\ee
The same result could have been easily obtained on using directly the expression of the basis elements $v^j_{m\,\tilde m}$ and of the coordinates variables as functions on $\R^4$ and using the Moyal product.
We have then
\be
x_0\star x_0\star v^j_{m\tilde m}= \lambda^2 ( j+\frac{1}{2})^2 \, v^j_{m\tilde m} \label{x0x0}
\ee
which is different from
\be
\left( \frac{1}{2}(x_-\star x_+ +x_+\star x_-)+  x_3\star x_3\right)v^j_{m\tilde m} = \lambda^2 j ( j+1)\, v^j_{m\tilde m} \label{x+x-}
\ee

\section{The Duflo map}\label{sec.3}

The Duflo isomorphism \cite{Duflo} first appeared in Lie theory and representation
theory. For a detailed and pedagogical  review we refer to \cite{onduflo}. It is an algebra
isomorphism between invariant polynomials of a Lie algebra and the center of its
universal enveloping algebra.
Kontsevich \cite{Kontsevich} later refined Duflo's result in the framework of
deformation quantization.

 Since the fundamental results by Harish-Chandra and others, it
is now well understood that the algebra of invariant polynomials on the dual of a Lie algebra of a particular type (solvable, synple or nilpotent)
 is
isomorphic to the center of the corresponding universal enveloping algebra. This fact
was generalized to an arbitrary finite-dimensional real Lie algebra by  Duflo in
\cite{Duflo}.  This isomorphism is called
the Duflo isomorphism  and it is obtained on composing  the  Poincar\'e-Birkhoff-Witt isomorphism (which is only an isomorphism at the level of vector spaces)
with an automorphism of the space of polynomials.

Because of its very definition the Duflo is a mathematically preferred quantization map, although it is often objected in the physics literature  that its application to physics is ambigous as it doesn't reproduce the correct energy spectrum for well known  problems in quantum mechanics like the quantization of the hydrogen atom.\footnote{ For applications to Quantum Gravity see \cite{Majid, noui2}.}  Hereafter we will first derive the map for the quadratic Casimir of the rotation group and then apply it precisely to the hydrogen atom showing that we obtain the correct spectrum if the whole $SO(4)$ invariance of the system is taken into account.

Let us indicate with $S(\mathfrak{g})$ the symmetric algebra over the Lie algebra
$\mathfrak{g}$. This may be identified with the polynomial functions of
$S(\mathfrak{g})^*$. This is isomorphic to the universal enveloping algebra
$U(\mathfrak{g})$ as a vector space. Let us consider the subalgebra of
 $\mathrm{ad}_{\mathfrak{g}}$ invariant polynomials, $S(\mathfrak{g})^\mathfrak{g} $
and  let us define the Poincar\'e-Birkhoff-Witt map
\be
I_{PBW}: S(\mathfrak{g})^\mathfrak{g}\rightarrow \mathcal{Z}(U(\mathfrak{g}))
\ee
\be
I_{PBW}(x_1,...x_n)= \frac{1}{n!}\sum_{\sigma(p)\in\Pi(n)} x_{\sigma(1)} ..x_{\sigma(n)}
\ee
with $\mathcal{Z}(U(\mathfrak{g}))$ the center of $U(\mathfrak{g})$.
This is an isomorphism of vector spaces but not in general an isomorphism of algebras.
Generalising previous results of Haris-Chandra to all finite dimensional Lie algebras
Duflo proved in \cite{Duflo} that it could be extended to an algebra isomorphism.
Upon defining
\be
j^{\frac{1}{2}}(x):= {\det}^{\frac{1}{2}}\left[\frac{\sinh\frac{1}{2} \mathrm{ad}\,
x}{\frac{1}{2} \mathrm{ad}\, x}\right]
\ee
the Duflo isomorphism
\be
\chi_D: S(\mathfrak{g})^{\mathfrak{g}}\rightarrow \mathcal{Z}( U({\mathfrak{g}}))
\label{duflo}
\ee
is proven to be
\be
\chi_D=I_{PBW}\circ j^{\frac{1}{2}}(\del)
\ee
where
\be
\del:= \tau_i\frac{\del}{\del x^i}
\ee
 $\tau_i$ are the generators of the Lie algebra in the adjoint representation and $\del/\del x^i$ are differential operators acting on the universal enveloping algebra. For
 the $\mathfrak{su}(2)$ case the adjoint representation is given by the defining representation of
 $\mathfrak{so}(3)$, that is
\be
 \tau_1=\left(
   \begin{array}{ccc}
     0 & 0 & 0 \\
     0 & 0 & -\ii \\
     0 & \ii & 0 \\
   \end{array}
 \right)
\;\;\; \tau_2=\left(
   \begin{array}{ccc}
     0 & 0 & \ii \\
     0 & 0 & 0 \\
     -\ii & 0 & 0 \\
   \end{array}
 \right)\;\;\;
 \tau_3=\left(
   \begin{array}{ccc}
     0 & -\ii & 0 \\
     \ii & 0 & 0 \\
     0 & 0 & 0 \\
   \end{array}
 \right)
\ee
Up to the second order in the generators we have then
\be
j^{\frac{1}{2}}(x)={\det}^\frac{1}{2}\left[\mathbf{1}+\frac{(\tau_i x_i)^2}{24} +
O(x^4)\right]= 1+ \frac{1}{2} \tr\frac{(\tau_i x_i)^2}{24}= 1+
\frac{1}{2}\frac{|x|^2}{12}= 1+ \frac{|x|^2}{24} \label{jexp}
\ee
where we have used
\be
\ln\det (\mathbf{1}+A)=\sum_1^\infty \frac{(-1)^{n+1}}{n!}\tr A^n
\ee
When replacing \eqn{jexp} in the Duflo isomorphism \eqn{duflo} and applying it to
polynomials of second order we obtain
\be
\chi_D (x_j x_k)= I_{PBW}\circ \left(1+\frac{\del_i\del_i}{24}\right)(x_j x_k)= I_{PBW}
(x_j x_k ) + \frac{1}{12} \delta_{jk}
\ee
When applied to the Casimir function  $x_j x_j$ we finally obtain
\be
I_{PBW} (x_j x_j)=   X_j  X_j  +  \frac{1}{4} \label{Duflosu2}
\ee
where $x_j x_j$ is an element in the symmetric algebra, that is the symmetric functions on $\R^3$ which will be our classical observables, while $X_j X_j$ indicates a polynomial in the center of the universal enveloping algebra. We then  represent the latter as an operator acting on  the  Hilbert space $\mathfrak{X}=\{ |j, m\rangle , j \in \mathbb{N}/2,\, -j\le m\le j\}$ of the angular momentum eigenstates
\be
\rho: X_j X_j \rightarrow \hat X_j \hat X_j : \mathfrak{X} \mapsto \mathfrak{X} \label{hilbertsu2}
\ee
so that the Duflo quantization prescription for the  symmetric function $x_j x_j$ reads
\be
\hat\chi:x_j x_j\rightarrow \hat X_j\hat X_j+\frac{1}{4}\mathbb{I} \label{su2qmap}
\ee
with $\hat \chi=\rho\circ\chi$
and we obtain for the spectrum
\be
\hat\chi(x_jx_j)|j m\rangle=(j(j+1)+\frac{1}{4})|j m\rangle=(j+\frac{1}{2})^2|j m\rangle
\ee
 Let us stress the factor $1/4$ which modifies the spectrum. In next section we shall apply this quantization scheme to the hydrogen atom.
 \subsection{The hydrogen atom}\label{hydrogen atom}
The hydrogen atom belongs to the well known class of Keplerian systems, its reduced Hamiltonian being of the form
\be
H= \frac{p^2}{2 m}-\frac{k }{r}
\ee
with  $m$ the reduced mass  of the electron and $k$ the Coulomb constant multiplied by the square of the electron charge. It is well known that the system is  super-integrable, with 7 constants of motion, 5 of which being independent. Besides the Hamiltonian, we have the angular momentum
\be
\mathbf{L}=\mathbf{r}\times \mathbf{p}
\ee
and the Runge-Lenz vector
\be
\mathbf{A}={\mathbf{p} \times \mathbf{L}}-{m} k\frac{\mathbf{r}}{r}
\ee
with
\be
\mathbf{A}\cdot \mathbf{L}=0,\;\;\; \mathbf{A}^2= {2}{m}H \mathbf{L^2}+m^2 k^2 \label{conditions}
 \ee
 On considering only negative values for the energy (as we will be interested in the discrete spectrum of  the quantum system) we define
\be
\mathbf{D}=\frac{\mathbf{A}}{\sqrt 2m |E|}
\ee
The angular momentum and rescaled Runge-Lenz components close the $\mathfrak{so}(4)$ Lie algebra w.r.t. the canonical Poisson brackets on $T^*\R^3$, $\{p_i,q_j\}=\delta_{ij}$.
 We have indeed
 \beqa
 \{L_i,L_j\}&=&\epsilon_{ij k}L_k\\
 \{D_i,D_j\}&=&\epsilon_{ij k}L_k \\
 \{L_i,D_j\}&=&\epsilon_{ij k}D_k
 \eeqa
 On introducing two sets of mutually commuting $\mathfrak{su}(2)$ generators
\be
 \mathbf{B}= \frac{1}{2}(\mathbf{D}+\mathbf{L}),\;\;\; \mathbf{C}= \frac{1}{2}(\mathbf{D}-\mathbf{L})
 \ee
we have
\be
\mathbf{D}^2+\mathbf{L^2}=2(\mathbf{B}^2+\mathbf{C}^2)=\frac{m k^2}{2 |E|} \label{energy}
\ee
that is, the energy can be expressed in terms of the two quadratic Casimir functions of the $\mathfrak{so}(4)\simeq\mathfrak{su}(2)\otimes \mathfrak{su}(2)$ algebra.

In order to achieve the quantization and obtain the energy spectrum for bounded states we resort to the Duflo quantization map  \eqn{su2qmap} and obtain
\be
\hat \chi(B_j B_j)= \hat B_j \hat B_j + \frac{1}{4}\mathbf{I} \;\;\; \hat\chi(C_j C_j)= \hat C_j \hat C_j + \frac{1}{4}\hat{\mathbf{I}}
\ee
so that
\be
\hat\chi(\mathbf{B}^2+\mathbf{C}^2)=\hat{\mathbf{B}}^2+\hat{\mathbf{C}}^2+\frac{1}{2}\hat {\mathbf{I}}
\ee
Resorting to the relation with the Runge-Lenz vector and the angular momentum we find
\be
\hat \chi(\mathbf{D}^2+\mathbf{L}^2)=\hat\chi\left[2(\mathbf{B}^2+\mathbf{C}^2)\right]=2(\hat{\mathbf{B}}^2+\hat{\mathbf{C}}^2)+\hat {\mathbf{I}}
\ee
We then observe that $\mathbf{B}^2= \mathbf{C}^2$ because of $\mathbf{D}\cdot\mathbf{L}=0$. Therefore
\be
\hat\chi( {\mathbf{D}}^2+\hat {\mathbf{L}}^2 )|j,m\rangle=(4 \hat{\mathbf{B}}^2 +\hat {\mathbf{I}})|j,m\rangle = (4j(j+1)+1) |j,m\rangle
\ee
Posing $2 j= n$ with $n$ integer and solving for the energy we finally find
\be
|E|=\frac{m k^2 }{2(n+1)^2}
\ee
which is the correct result.

\section{The hydrogen atom in the $\star$-quantization approach. Conclusions}\label{sec.4}

 In order to compare with the $\star$-quantization approach  induced by the star product we have derived in section \ref{sec.2}
 we have explicitly computed the action of the $\mathfrak{su}(2)$ Lie
algebra generators in the matrix basis in section \ref{angularmomentum}.
We have found for the Casimir function $x_0$
\be
x_0\star v^j_{m\tilde m} =  v^j_{m\tilde m} \star x_0= \lambda ( j+\frac{1}{2}) \,v^j_{m\tilde m}
\ee
so that
\be
x_0\star x_0\star v^j_{m\tilde m}= v^j_{m\tilde m}\star x_0\star x_0=\lambda^2 ( j+\frac{1}{2})^2 \, v^j_{m\tilde m}
\ee
This is in fact the symbol of the eigenvalue equation
\be
\hat { \mathbf{X}}^2 |j, m\rangle=c(j,m) |j,m\rangle
\ee
in agreement with the result of the Duflo quantization as in \eqn{Duflosu2}. Let us notice that, had we chosen a different $\star$-product on $\R^4$ than the Moyal product,   compatible with the same $\star$-commutation relations for the coordinate functions \eqn{starcommz}, we would have obtained a different result for the quantization of the Casimir function, although inducing the same $\star$-commutation relations for the $\R^3$ coordinate functions as in \eqn{inducedstarcom}. For example, the Voros product, in the properly defined matrix basis, $\tilde v^j_{m\tilde m}$   would give \cite{vitalewallet}
\be
x_0\star_V \tilde v^j_{m\tilde m} =  \tilde v^j_{m\tilde m} \star_V x_0= \lambda  j \,\tilde v^j_{m\tilde m}
\ee
and the constant shift is not at all anodyne when the result is applied to the hydrogen atom spectrum. This is probably due to the fact that only the Moyal quantization is selfdual (meaning that we can use the same operator to map the observables and the states to noncommutative symbols). This suggests that in other quantization schemes we shouldn't use the same matrix basis  for the observables and the states as indicated in \cite{conVolodya} (also see  \cite{scholz} were the problem of the equivalence of Moyal and Voros quantization is discussed). We shall come back to this issue elsewere.

To conclude, the Moyal-induced $\star$ product on $\R^3$ first introduced in \cite{selene} is singled out among other $\star$-quantization schemes in that it  agrees with the Duflo quantization map of $\mathfrak{su}(2)$ Casimir elements. We speculate that the same result will old true for Casimir functions of other three dimensional Lie algebras, whose corresponding Moyal-induced $\star$ products on $\R^3$ are illustrated in \cite{selene}.
Moreover we stress once again that the correct spectrum of quantum systems with a central potential is reproduced in the quantization scheme discussed in the paper.

\section*{Acknowledgements}
P. Vitale thanks the Laboratoire de Physique Th\'eorique at  Orsay for hospitality while this paper was partially written and acknowledges a grant from the
European Science Foundation under the research networking project  ”Quantum Geometry and
Quantum Gravity”, and partial support by GDRE GREFI GENCO. She also thanks Harald Grosse and Jean-Christophe Wallet for enlightening discussions.


\begin{thebibliography}{99}
\bibitem{qgpeople}
  L.~Freidel and E.~R.~Livine,
  ``Effective 3-D quantum gravity and non-commutative quantum field theory,''
  \emph{Phys.\ Rev.\ Lett.} {\bf 96}, 221301 (2006)
  [arXiv:hep-th/0512113].
    A.~Baratin and D.~Oriti,
  ``Group field theory with non-commutative metric variables,''
  \emph{Phys.\ Rev.\ Lett.} {\bf 105}, 221302 (2010)
  [arXiv:1002.4723 [hep-th]].


 \bibitem{Majid} L.~Freidel and S.~Majid,
  ``Noncommutative harmonic analysis, sampling theory and the Duflo map in 2+1 quantum gravity,''
  \emph{Class.\ Quant.\ Grav.} {\bf 25}, 045006 (2008)
  [arXiv:hep-th/0601004].

 \bibitem{noui}
  E.~Joung, J.~Mourad and K.~Noui,
  {\it Three Dimensional Quantum Geometry and Deformed Poincare Symmetry,}
  \emph{J.\ Math.\ Phys.} {\bf 50}, 052503 (2009)
  [arXiv:0806.4121 [hep-th]].


\bibitem{vitalewallet} Patrizia Vitale, Jean-Christophe Wallet ``Noncommutative field theories on $\mathbb{R}_\lambda^3$ and matrix models'' Orsay preprint LPT-12-97.

\bibitem{PV12} Patrizia Vitale ''Aspects of group field theory'' AIP Conference Proceedings {\bf 1460} (2012) 101

\bibitem{selene}
J.~M.~Gracia-Bond\'ia, F.~Lizzi, G.~Marmo and P.~Vitale,
  ``Infinitely many star products to play with,''
  \emph{JHEP} {\bf 0204}, 026 (2002)
  [arXiv:hep-th/0112092];

\bibitem{pepejoe} J. M. Gracia-Bond\'ia and J. C. V\'arilly, ``Algebras Of Distributions Suitable
For Phase Space Quantum Mechanics. 1'', J. Math. Phys. 29 (1988) 869.


\bibitem{grossewulkenhaar} H. Grosse and R. Wulkenhaar, ``Renormalisation of $\phi^4$ theory on noncommutative
$\mathbb{R}^2$
in the matrix base,'' JHEP 0312 (2003) 019
 H.~Grosse and R.~Wulkenhaar,
  ``Renormalization of phi**4 theory on noncommutative R**4 in the matrix base,''
  Commun.\ Math.\ Phys.\  {\bf 256} (2005) 305
  [hep-th/0401128].

\bibitem{Gronewold} H. Gr\"onewold, ``On the Principles of Quantum
    Mechanics'', Physica {\bf12} (1946)~405.

\bibitem{Moyal} J.~E.~Moyal,
  ``Quantum Mechanics as a Statistical Theory,''
  Proc.\ Cambridge Phil.\ Soc.\  {\bf 45} (1949)~99.

 \bibitem{Weyl} H. Weyl The theory of Groups and
Quantum Mechanics Dover (1931), translation of Gruppentheorie und Quantemmechanik,
Hirzel Verlag (1928).

\bibitem{Wigner} E.P. Wigner, ``On the Quantum Correction for
Thermodynamic Equilibrium'', Phys. Rev. 40 (1932) 749.

\bibitem{conVolodya} V. I. Man'ko, G. Marmo and P.
Vitale, ``Duality symmetry for star products,'' Phys. Lett. A 334 (2005) 1.
[arXiv:hep-th/0407131].\\
 O. V. Man'ko, V. I. Man'ko, G. Marmo and P. Vitale, ``Star
products, duality and double Lie algebras,''  Phys. Lett. A 360 (2007) 522.
[arXiv:quant-ph/0609041].

\bibitem{bayen} F. Bayen, M. Flato, C. Fronsdal, A. Lichnerowicz and D. Sternheimer, ``Deformation
theory and quantization. I. Deformation of symplectic structures'', Ann. Phys. (NY) 111
(1978) 61; ``Deformation theory and quantization. II. Physical applications'', Ann. Phys.
(NY) 111 (1978) 111.

\bibitem{zampinithesis} A. Zampini, ``Applications of the Weyl-Wigner
formalism to noncommutative geometry,''  arXiv:hep-th/0505271.


\bibitem{VarillyGraciaBondia}R. Estrada, J. M. Gracia-Bondia and J. C. Varilly, ``On Asymptotic expansions of
twisted products,'' J. Math. Phys. 30 (1989) 2789.

\bibitem{voros} A. Voros, ``Wentzel-Kramers-Brillouin method in the Bargmann representation'',
Phys. Rev. A {\bf 40} 6814 (1989).

\bibitem{GGISV04} V. Gayral, J. M. Gracia-Bondia, B Iochum, T. Sch\"ucker and J. C. Varilly, Moyal
planes are spectral triples, Comm. Math. Phys. 246 (2004) 569.

\bibitem{MMVZ94}  V.~I.~Manko, G.~Marmo, P.~Vitale and F.~Zaccaria,
  ``A Generalization of the Jordan-Schwinger map: Classical version and its q deformation,''
  Int.\ J.\ Mod.\ Phys.\ A {\bf 9}, 5541 (1994)
  [hep-th/9310053].

\bibitem{HLS}
 A.~B.~Hammou, M.~Lagraa and M.~M.~Sheikh-Jabbari,
  ``Coherent state induced star product on R**3(lambda) and the fuzzy sphere,''
  {Phys.\ Rev.\ D}  {\bf 66}, 025025 (2002)
  [arXiv:hep-th/0110291].

\bibitem{GLV08} S.~Galluccio, F.~Lizzi and P.~Vitale,
  ``Twisted Noncommutative Field Theory with the Wick-Voros and Moyal Products,''
  Phys.\ Rev.\ D {\bf 78} (2008) 085007
  [arXiv:0810.2095 [hep-th]].

\bibitem{GLV09} S.~Galluccio, F.~Lizzi and P.~Vitale,
  ``Translation Invariance, Commutation Relations and Ultraviolet/Infrared Mixing,''
  JHEP {\bf 0909} (2009) 054
  [arXiv:0907.3640 [hep-th]].

\bibitem{TV10} A. Tanasa and P. Vitale, A.~Tanasa and P.~Vitale,
  ``Curing the UV/IR mixing for field theories with translation-invariant $\star$ products,''
  Phys.\ Rev.\ D {\bf 81}, 065008 (2010)
  [arXiv:0912.0200 [hep-th]].


\bibitem{pepespin}  J. M. Gracia-Bond\'ia and J. C. V\'arilly, ``The Moyal representation for spin'' Annals of Physics {\bf 190}, 107 (1989) \\
H Figueroa, J. M. Gracia-Bond\'ia and J. C. V\'arilly, ``Moyal quantization with compact symmetry groups and noncommutative harmonic analysis''
  Journal of Mathematical Physics {\bf 31}, 2664 (1990)

\bibitem{volodyabeppe}
  O.~V.~Man'ko, V.~I.~Man'ko and G.~Marmo,
  ``Star product of generalized Wigner-Weyl symbols on SU(2) group, deformations, and tomographic probability distribution,''
  Phys.\ Scripta {\bf 62}, 446 (2000).

\bibitem{MMT} S.~Mancini, V.~I.~Man'ko and P.~Tombesi,
  ``Symplectic tomography as classical approach to quantum systems,''
  Phys.\ Lett.\ A {\bf 213}, 1 (1996)
  [quant-ph/9603002].

\bibitem{noiconvolodya}
  V.~I.~Manko, L.~Rosa and P.~Vitale,
  ``Probability representation in quantum field theory,''
  Phys.\ Lett.\ B {\bf 439}, 328 (1998)
  [hep-th/9806164].
  \\
  V.~I.~Man'ko, L.~Rosa and P.~Vitale,
  ``Time dependent invariants and Green functions in the probability representation of quantum mechanics,''
  Phys.\ Rev.\ A {\bf 57}, 3291 (1998)
  [quant-ph/9802030].

\bibitem{diffcalc} G. Marmo, P. Vitale and A. Zampini  ``Noncommutative differential calculus for Moyal subalgebras'' J. Geom. Phys. {\bf 56} (2006) 611
[arXiv:hep-th/0411223]

\bibitem{discofuzzy}  F.~Lizzi, P.~Vitale and A.~Zampini,
  ``The Fuzzy disc,''
  JHEP {\bf 0308} (2003) 057
  [hep-th/0306247].
    ``The Beat of a fuzzy drum: Fuzzy Bessel functions for the disc,''
  JHEP {\bf 0509} (2005) 080
  [hep-th/0506008].
    ``The fuzzy disc: A review,''
  J.\ Phys.\ Conf.\ Ser.\  {\bf 53}, 830 (2006).
    ``From the fuzzy disc to edge currents in Chern-Simons theory,''
  Mod.\ Phys.\ Lett.\ A {\bf 18}, 2381 (2003)
  [hep-th/0309128].

\bibitem{Duflo} M. Duflo, Op\'erateurs diff\'erentiels bi-invariants sur un groupe de Lie.
Ann. Sc.Ec. Norm. Sup. 10 (1977), 107-144.

\bibitem{onduflo} Damien Calaque, Carlo A. Rossi "Lectures on Duflo Isomorphisms in Lie Algebra and Complex Geometry"
(EMS Series of Lectures in Mathematics) (2011)

\bibitem{Kontsevich} M. Kontsevich, ``Deformation quantization of Poisson manifolds, I'',  Lett. Math. Phys.
66 (2003) 157. [arXiv:q-alg/9709040].

\bibitem{noui2}
  K.~Noui, A.~Perez and D.~Pranzetti,
  ``Canonical quantization of non-commutative holonomies in 2+1 loop quantum gravity,''
  JHEP\ {\bf 1110} (2011) 036
  [arXiv:1105.0439 [gr-qc]].

\bibitem{scholz} \bibitem{Basu:2011kh}
  P.~Basu, B.~Chakraborty and F.~G.~Scholtz,
  ``A Unifying perspective on the Moyal and Voros products and their physical meanings,''
  J.\ Phys.\ A A {\bf 44}, 285204 (2011)
  [arXiv:1101.2495 [hep-th]].

\end{thebibliography}
\end{document}